\documentclass[prd,10pt,aps,showpacs,amsmath,amssymb,twocolumn,nofootinbib,longbibliography]{revtex4-1}

\usepackage{natbib}
\usepackage[colorlinks]{hyperref}
\hypersetup{linkcolor=blue,citecolor=blue}

\usepackage{graphicx}
\usepackage{dcolumn}
\usepackage{bm}
\usepackage{color}
\usepackage{soul}
\renewcommand{\thefootnote}{\fnsymbol{footnote}}

\newcommand{\mnras}{MNRAS} 
\usepackage{hyperref}

\def\apj{ApJ}
\def\mnras{MNRAS}
\def\apjl{ApJL}

\usepackage{color}
\usepackage{soul}
\usepackage{graphicx}
\usepackage{dcolumn}
\usepackage{bm}

\begin{document}


\title{Born-Infeld magnetars: larger production of gravitational waves due to larger toroidal magnetic fields}

\author{Jonas P. Pereira$^\ast$}
\affiliation{Universidade Federal do ABC, Centro de Ci\^encias Naturais e Humanas, Avenida dos Estados 5001, 09210-170, Santo Andr\'e, SP, Brazil}
\email{jonas.pereira@ufabc.edu.br}

\author{Jaziel G. Coelho$^\dagger$}
\affiliation{%
Departamento de F\'isica, Universidade Tecnol\'ogica Federal do Paran\'a, 85884-000 Medianeira, PR, Brazil
}%
\email{jazielcoelho@utfpr.edu.br}
\author{Rafael C. R. de Lima}
\affiliation{%
 Universidade do Estado de Santa Catarina, Av. Madre Benvenuta, 2007 Itacorubi, 88.035-901, Florian\'opolis, Brazil
}%

\date{\today}

\begin{abstract}
We discuss some aspects of \citet{2018EPJC...78..361P} concerning magnetars described by nonlinear theories of the electromagnetism and make the case for the Born-Infeld Lagrangian. We focus on the increase of toroidal magnetic fields in these systems with respect to ordinary magnetars and the subsequent increase of gravitational wave production. In summary, nonlinear theories of the electromagnetism would make it more likely for the detection of gravitational waves with future detectors, which could constrain nonlinear aspects of electrodynamics not entirely possible on Earth-based particle accelerators.
\end{abstract}
\maketitle

\section{Introduction}

Although magnetar phenomenology is relatively well-known \citep{2017ARA&A..55..261K}, we are far from fully understanding their physics. They have been conceived as systems with supercritical surface magnetic fields ($B>4\times 10^{13}$ G and usually of order $10^{14}-10^{15}$ G) in order to explain their burst, outburst and flare activities \citep{1992ApJ...392L...9D,1995MNRAS.275..255T,2017ARA&A..55..261K}. However, subtleties are also present in these systems. For instance, there are observationally known cases where they could have subcritical dipolar magnetic fields \citep{2010Sci...330..944R,2012ApJ...754...27R,2014ApJ...781L..17R}, 
which naturally drew interest to new models with predominant toroidal magnetic fields \citep{2013MNRAS.435L..43C}. Therefore, it is now expected that magnetars have huge (hidden) toroidal magnetic fields driving their phenomenology, and this clearly motivates analysis of these configurations within nonlinear electrodynamics. Apart from clear evidence thereof in the context of QED \citep{2010PhR...487....1R}, there are many other candidates for nonlinear electrodynamics. We focus here on an older proposal due to Born and Infeld \citep{1934RSPSA.144..425B}, known today to be a low energy limit to string theory \citep{1997hep.th....2087R}. The motivation for this comes from the ``gap'' left in the probe of its scale field $b$ in some ATLAS experiments \citep{2017PhRvL.118z1802E} and the fact it could exactly be ``filled up'' by expected fields in magnetars \citep{2018EPJC...78..361P}. Though it seems unlikely the scale field to the Born-Infeld Lagrangian should lie in such ``small'' field range when string theory is taken into account, for physically meaningful constraints it is important not to leave any range of $b$ behind, and it is very serendipitous that what cannot be probed on Earth could exactly be done with magnetars.

\section{Nonlinear electrodynamics within ideal MHD}

Stars are believed to be highly ionized systems, which should render them good conductors. We assume this to be the case here and take the limit they are ideal conductors, allowing us to work with ideal magnetohydrodynamics (MHD) \citep{2009lema.book.....S}. The relevant field equations in the presence of nonlinear electrodynamics are (we work with Gaussian units)
\begin{equation}
\nabla\cdot (L_F\vec{E})=4\pi \rho\label{divE},
\end{equation}
\begin{equation}
\nabla\cdot \vec{B}=0\label{divB},
\end{equation}
\begin{equation}
\nabla\times \vec{E}=-\frac{1}{c}\frac{\partial \vec{B}}{\partial t}\label{rotE}
\end{equation}
and
\begin{equation}
\nabla\times (L_F\vec{B})=\frac{4\pi}{c} \vec{j}\label{rotB},
\end{equation}
where $L_F\equiv \partial L/\partial F$, $F\equiv F^{\mu\nu}F_{\mu\nu}$ ($F_{\mu\nu}$ is the electromagnetic tensor), and $L$ is the Lagrangian density of the electromagnetism \citep{2018EPJC...78..361P}. Besides, we have defined $\rho$ as the system's charge density and $\vec{j}$ its current vector. For a good conductive region of a rigidly rotating star with an angular frequency $\vec{\omega}$ and small tangential velocities $\vec{v}$ ($v/c\ll 1$ $v\doteq ||\vec{v}||$), it follows that
\begin{equation}
\vec{E}=-\frac{\vec{v}}{c}\times\vec{B}=-\frac{\vec{\omega}\times \vec{r}}{c}\times \vec{B}\label{solEconduct},
\end{equation}
which comes from the assumption that Ohm's law also holds for nonlinear electrodynamics, assumed here. 
One can eliminate $\rho$ and $\vec{j}$ from Eqs. \eqref{divE} and \eqref{rotB} by means of Maxwellian (Ma) fields, and is left with
\begin{equation}
L_F\vec{E}=\vec{E}_{Ma}+\nabla \times \vec{C}\label{solE}
\end{equation}
and
\begin{equation}
L_F\vec{B}=\vec{B}_{Ma} + \nabla f\label{solB},
\end{equation}
where $\vec{C}$ and $f$ are arbitrary functions. Consistency with MHD [see Eq. \eqref{solEconduct}] implies that
\begin{equation}
\nabla \times \vec{C}= -\frac{\vec{\omega}\times \vec{r}}{c}\times \nabla f\label{C-f-relationship}.
\end{equation}
For the majority of cases, Eqs. \eqref{solE} and \eqref{solB} are merely indicative, due to the difficulty in finding $\vec{C}$, $f$, or even the Maxwellian fields. However, for the case of toroidal fields, the equations above greatly simplify the problem.

\section{Toroidal Born-Infeld fields}

We constrain ourselves to the case of dominating axisymmetric toroidal fields, where $\vec{B}=B_{\phi}(r,\theta)\hat{\phi}$ ($\hat{\phi}$ is the azimuthal unit vector). This field is such that Eq. (\ref{divB}) is automatically satisfied. From Eq.~(\ref{solEconduct}), toroidal fields in rigidly rotating stars (we define the $z$-axis such that $\vec{\omega}=\omega \hat z$, which means $\vec{v}=v\hat \phi$, $v$ being any) result in $\vec{E}=\vec{0}$, which automatically satisfies Eq. (\ref{rotE}) and from Eqs.~(\ref{solE}) and (\ref{C-f-relationship}) leads to $\nabla f = \vec{0}$. Thus, from Eq.~(\ref{solB}) (the only remaining nonlinear equation)
\begin{equation}
L_FB_{\phi}= B^{Ma}_{\phi}\label{Btoroidal}.
\end{equation}
Given that in the small field regime of nonlinear electrodynamics $L_F=1-|\mbox{something small}|$ \citep{2016ApJ...823...97C}, one generically has from the above equation that nonlinear toroidal fields in ideal MHD are larger than their Maxwellian counterparts.

Let us particularize the above to the case of the Born-Infeld Lagrangian taking into account Eq. \eqref{solEconduct}. 
It is defined as  \citep{2018EPJC...78..361P}
\begin{equation}
L_{B.I}\doteq 4b^2\left(\sqrt{1+\frac{F}{2b^2}}-1\right)\label{LB-I},
\end{equation}
where $b$ is the scale field to the theory and $F= 2(\vec{B}^2- \vec{E}^2)\approx B^2$. From Eqs. (\ref{Btoroidal}) and (\ref{LB-I}) one has that
\begin{equation}
{B}_{\phi}=\frac{b{B}^{Ma}_{\phi}}{\sqrt[]{b^2-(B^{Ma}_{\phi})^2}}\label{BBItoroidal},
\end{equation}
which clearly shows that ${B}_{\phi}>{B}^{Ma}_{\phi}$. Besides, for the energy density $\rho$ of the electromagnetic fields \citep{2018EPJC...78..361P}
\begin{equation}
\frac{\rho_{BI}}{\rho_{Ma}}= 2\left(\frac{b}{B^{Ma}_{\phi}} \right)^2 \left\{\left[1-\left(\frac{B_{\phi}^{Ma}}{b} \right)^2\right]^{-\frac{1}{2}} -1 \right\},\label{energy_ratio_BI}
\end{equation}
which is always larger than unit for $B_{\phi}^{Ma}<b$, needed for the consistency of Born-Infeld toroidal fields, as clearly expressed by Eq. \eqref{BBItoroidal}.

\section{Some consequences}

Larger magnetic fields should deform more stars, which would imply larger magnetic ellipticities $\epsilon$ and hence production of gravitational waves. Indeed, when $\epsilon$ is proportional to the total magnetic field squared, it follows that \citep{2018EPJC...78..361P}
\begin{equation}
\frac{\epsilon_{BI}}{\epsilon_{Ma}}= \left(\frac{\dot{E}_{GW}^{BI}}{\dot{E}_{GW}^{Ma}}\right)^{\frac{1}{2}}\approx \left(\frac{\Delta{E}_{GW}^{BI}}{\Delta{E}_{GW}^{Ma}}\right)^{\frac{1}{2}}\approx \left [ 1-\left(\frac{B_{\phi}^{Ma}}{b} \right)^2 \right]^{-1}\label{GW},
\end{equation}
where $\dot{E}_{GW}$ is the energy loss due to gravitational waves. It has been assumed that $B_{\phi}^{Ma}$ is slowly varying within the star, which is reasonable for confined fields and maximal estimates. Given that $B_{\phi}^{Ma}<b$ [see Eq. (\ref{BBItoroidal})], one has that the production of gravitational waves is larger in the Born-Infeld theory than in Maxwell's, as expected. 

From Eq. \eqref{GW}, given a lower limit to $b$ one could find an upper limit to $\epsilon_{BI}$, $\epsilon_{ul}$,  which is
\begin{equation}
|\epsilon_{ul}|=\left[1-\left(\frac{B_{\phi}^{Ma}}{b_{min}}\right)^2 \right]^{-1}|\epsilon_{Ma}|.\label{epsilonulbmin}
\end{equation}
At the same time, an upper limit to $\epsilon_{BI}$ would imply an absolutely minimum value for $b$. The importance of the above is that gravitational wave observations can already constrain ellipticities, which could thus constrain aspects to scale fields from nonlinear theories. Let us make some estimates for the Born-Infeld Lagrangian. 

Averaged toroidal Maxwellian fields could be \textit{estimated} from giant flare events with the help of
\begin{equation}
(B_{\phi}^{Ma})^2= \frac{6 E_{fl}}{R^3}\label{Bphi-mean},
\end{equation}
with $E_{fl}$ the energy released during a magnetar flare event and $R$ the stellar radius. From it, one could easily estimate the Maxwellian magnetic ellipticity through \citep{2018EPJC...78..361P} $|\epsilon_{Ma}|= R^4/(G M^2)(B_{\phi}^{Ma})^2= 6 R E_{fl}/(G M^2)$, where $M$ is the magnetar's mass. Taking $E_{fl}=10^{47}$ erg, related to the energy budget of the largest flare of magnetar SGR 1806--20 \citep{2017ARA&A..55..261K}, and by making use of fiducial magnetar parameters [$R=10$ km and $M=1.4$ M$_{\bigodot}$], it follows that  
$|\epsilon_{Ma}|= 1.20\times 10^{-6}$. The absolutely minimum value of $b$ should be the one determined by Born-Infeld themselves by means of the electron properties, namely $b_{min}\approx 3.96 \times 10^{15}$ G. Thus, from the above, $|\epsilon_{ul}|\approx 1.24\times 10^{-6}$ and $\Delta{E}^{BI}_{GW}/\Delta{E}^{Ma}_{GW}\lesssim 1.08$.

Let us investigate other scenarios. Take now $E_{fl}=10^{47}-10^{48}$ erg, say $E_{fl}=5\times 10^{47}$ erg, which might be a possibility to giant flare events when nonlinear electrodynamics is present, or even to smaller than usual short GRBs [typical isotropic energy of ordinary short-GRBs are in the range $10^{49}-10^{52}$ erg \citep{2015JHEAp...7...73D}]. In this case, $|\epsilon_{ul}|\approx 7.4\times 10^{-6}$ [$B_{\phi}^{Ma}=1.7\times 10^{15}$ G] and $\Delta{E}^{BI}_{GW}/\Delta{E}^{Ma}_{GW}\lesssim 1.53$. Assume now magnetic fields just as high as $10^{15}$ G are possible in magnetars. From Eq. (\ref{GW}) it follows that $\Delta{E}^{BI}_{GW}/\Delta{E}^{Ma}_{GW}\lesssim 1.14$, which in turn results in $|\epsilon_{ul}|\approx 2\times 10^{-6}$.

\section{Conclusions}

We have shown that expected upper limits to the norm of the magnetic ellipticities of magnetars would be around $10^{-6}$ ($10^{-5}$) and the maximum increase of gravitational-wave energy released in Born-Infeld theory with respect to Maxwell's theory would be around $15\%$ (even larger than $50\%$) for flare energy up to $10^{47}$ ($10^{48}$) erg. From the above, naturally the best candidates for testing nonlinear electrodynamics are magnetars with large flare events, which would also imply much larger production of gravitational waves. No continuous gravitational wave detection from these systems would place upper limits to the ellipticity and hence lower limits to $b$, which could be cross-checked with already existing hydrogen atom experiments. The analysis presented here should just be seen as first order estimates for nonlinear effects in magnetars. Realistic systems should present both poloidal and toroidal fields and within nonlinear electrodynamics their study become much more intricate, due to naturally occurring couplings of the fields components. However, nonlinearities also add up terms to the equations, which could be important to address some problems in neutron stars such as known magnetic field instabilites \citep{2011MNRAS.412.1394L,2011MNRAS.412.1730L}. 

Summing up, we have stressed that toroidal fields in higly conductive nonlinear magnerars are larger than their Maxwellian counterparts. This increases the emission of gravitational waves from these stars. Current constraints on the Born-Infeld theory, giant-flare energetics and magnetic fields in magnetars suggest a maximum increase of $10\%-20\%$ of gravitational wave energy emitted from Born-Infeld magnetars. For larger giant-flare energies, much higher percentages may appear. Therefore, the possibility may present to probing nonlinear electrodynamics with the advancement of gravitational-wave detectors.

\section{ACKNOWLEDGEMENTS}
J.P.P. acknowledges the financial support given by Funda\c c\~ao de Amparo \`a Pesquisa do Estado de S\~ao Paulo (FAPESP) under grants No. 2015/04174-9 and 2017/21384-2. J.G.C. is likewise grateful to the support of Conselho Nacional de Desenvolvimento Cient\'ifico e Tecnol\'ogico - CNPq (421265/2018-3 and 305369/2018-0). R.C.R.L. acknowledges the support of Funda\c c\~ao de Amparo \`a Pesquisa e Inova\c c\~ ao do Estado de Santa Catarina (FAPESC) under grant No. 2017TR1761.

\nocite{*}

\bibliography{apssamp}

\end{document}